\documentclass[12pt]{iopart}

\usepackage{lineno,hyperref}
\usepackage{iopams}
\usepackage{graphicx}  
\begin{document}

\title[Fourier transformations and electromagnetic knots]{The method of Fourier transforms applied to electromagnetic knots}

\author{M Array\'as and J L Trueba}

\address{\'Area de Electromagnetismo, Universidad Rey Juan Carlos, Tulip\'
an s/n, 28933 M\'ostoles, Madrid, Spain}
\ead{manuel.arrayas@urjc.es, joseluis.trueba@urjc.es}
\vspace{10pt}

\begin{abstract}
The Fourier transform method can be applied to obtain electromagnetic knots, which are solutions of Maxwell equations in vacuum with non-trivial topology of the field lines and especial properties. The program followed in this work allows to present the main ideas and the explicit calculations at undergraduate level, so they are not obscured by a more involved formulation. We make use of helicity basis for calculating the electromagnetic helicity and the photon content of the fields.   
\end{abstract}

\section{Introduction}
The application of topology ideas to electromagnetism has led to the finding of new solutions of Maxwell equations in vacuum \cite{Arr17}, some of them with the property of non-nullity \cite{Arr15} and with other interesting properties in terms of helicity exchange \cite{Arr12} and spin-orbital decomposition \cite{Arr18}.

In this article we want to introduce one of the methods used in the current research of electromagnetic knots at undergraduate level. Only a basic knowledge of the Fourier transform is required by the student. The content of this article can be proposed to the students as a set of advanced problems exploring special topics such as the duality \cite{Stratton} and gauge transforms, the formulation of Cauchy problem in electromagnetism, helicity basis \cite{Ynd}, and photon polarization.   

We hope that it will help to clarify and review some concepts and ideas through explicit examples. For instance, in many introductory courses, electromagnetic fields in vacuum are presented in terms of planar waves. The planar electromagnetic waves are null fields \cite{Kosyakov}, and this leads quite often to the misconception that all the electromagnetic fields in vacuum must satisfy the property $|{\bf E}| = c|{\bf B}|,\, {\bf E}\cdot {\bf B}=0$. Here we work it out explicitly the calculations of non-null torus fields with the only use of Fourier transforms, so we obtain a solution where electric and magnetic fields are not orthogonal. At the same time the student is calculating a rather non-trivial solution of Maxwell equations in vacuum, which many of its properties being the subject of undergoing research \cite{Arr17a, Arr17b}.

We make a brief summary of Maxwell theory in vacuum, formulate the Cauchy problem and introduce the Fourier decomposition of the initial values of the fields. We show how solve the time evolution by making the inverse Fourier transform so expressing the fields as wave packages determined by the initial conditions. We then obtain the potential associated to the fields in the Coulomb gauge.

The helicity basis are used to express the potential as combinations of circularly polarized waves. The helicity basis are useful to introduce the magnetic, electric and electromagnetic helicities. The helicity of a vector field is a useful quantity to characterize some topological properties of the fields, such as the linkage of the field lines and the electromagnetic helicity can be related to the different of right- and left-photons content of the field in Quantum Mechanics.

We will apply this program to a set of knotted electromagnetic fields, including the particular case of the celebrated Hoft-Ra\~nada solution, called the Hopfion. 

\section{Maxwell theory in vacuum}
Electromagnetism in vacuum three-dimensional space can be described in terms of two real vector fields, ${\bf E}$ and ${\bf B}$, called the electric and magnetic fields respectively. Using the SI of Units, these fields satisfy the following Maxwell equations in vacuum,
\begin{eqnarray}
\nabla \cdot {\bf B} =0&,& \, \, \nabla \times {\bf E} + \frac{\partial
{\bf B}}{\partial t} =0, \label{elmaghel1} \\
\nabla \cdot {\bf E} =0&,& \, \, \nabla \times {\bf B} - \frac{1}{c^2} \frac{\partial
{\bf E}}{\partial t} =0, \label{elmaghel2}
\end{eqnarray}
where $c$ is the speed of light, $t$ is time and $\nabla$ denotes differentiation with respect to space coordinates ${\bf r} = \left( x, y, z \right)$.
The first two equations (\ref{elmaghel1}) are identities using the following four-vector electromagnetic potential in the Minkowski spacetime with coordinates $\left( x^{0} = c t, x^{1} = x, x^{2} = y, x^{3} = z \right)$ and metric $g = \mbox{diag} \, \left( 1 , -1, -1, -1 \right)$,
\begin{equation}
A^{\mu} = \left( \frac{V}{c} , {\bf A} \right).
\label{elmaghelpot} 
\end{equation}
In equation (\ref{elmaghelpot}), $A^{0} = V/c$, where $V$ is the scalar electric potential, and ${\bf A} = \left( A^{1}, A^{2}, A^{3} \right)$ is the vector electromagnetic potential. Greek indices as $\mu$ take values $0, 1, 2, 3$ and latin indices as $i$ take values $1, 2, 3$. An electromagnetic field tensor $F_{\mu \nu}$ can be defined as
\begin{equation}
F_{\mu \nu} = \partial_{\mu} A_{\nu} - \partial_{\nu} A_{\mu} ,
\label{elmaghelfield}
\end{equation}
with components related to the electric and magnetic fields through 
\begin{eqnarray}
{\bf E}_{i} &=& c \, F^{i0} , \nonumber \\
{\bf B}_{i} &=& - \frac{1}{2} \epsilon_{ijk} F^{jk} .
\label{elmaghelfield1}
\end{eqnarray}
In three-dimensional quantities, equation (\ref{elmaghelfield}) is
\begin{eqnarray}
{\bf E} &=& - \nabla V - \frac{\partial {\bf A}}{\partial t} , \nonumber \\
{\bf B} &=& \nabla \times {\bf A} .
\label{elmaghelpot1}
\end{eqnarray}
The first pair (\ref{elmaghel1}) of Maxwell equations become a pair of identities by using (\ref{elmaghelpot1}). The dynamics of electromagnetism in vacuum is then given by the second pair (\ref{elmaghel2}) of Maxwell equations, that can be written in spacetime quantities as
\begin{equation}
\partial _{\mu} F^{\mu \nu} = 0 .
\label{elmaghel4}
\end{equation}

Maxwell equations (\ref{elmaghel1})-(\ref{elmaghel2}) are invariant under the map $({\bf E} , c{\bf B} ) \mapsto (c{\bf B} , -{\bf E} )$, as stated in \cite{Stratton}. This means that, in vacuum, it is possible to define another four-potential
\begin{equation}
C^{\mu} = (c \, V^{\prime}, {\bf C}),
\label{cdefinition}
\end{equation}
so that the dual of the electromagnetic tensor $F_{\mu \nu}$, defined as
\begin{equation}
{}^{*}\!F_{\mu \nu} = \frac{1}{2} \varepsilon_{\mu \nu \alpha \beta} F^{\alpha \beta} ,
\label{elmaghel5}
\end{equation}
satisfies
\begin{equation}
{}^{*}\!F_{\mu \nu} =-\frac{1}{c} \left( \partial_{\mu} C_{\nu} - \partial_{\nu} C_{\mu} \right) ,
\label{elmaghel7}
\end{equation}
or, in terms of three-dimensional fields,
\begin{equation}
{\bf E} = \nabla \times {\bf C} , \, \, \, {\bf B} = \nabla V^{\prime} + \frac{1}{c^2} \frac{\partial {\bf C}}{\partial t} .
\label{elmaghel8}
\end{equation}
In this dual formulation, the pair of equations (\ref{elmaghel2}) are simply identities when definitions (\ref{elmaghel8}) are imposed, and the dynamics of electromagnetism in vacuum is given by the first pair of Maxwell equation (\ref{elmaghel1}), written in terms of components of the four-potential $C^{\mu}$.

Alternatively, Maxwell equations in vacuum can be described in terms of two sets of vector potentials, as in definitions (\ref{elmaghelfield}) and (\ref{elmaghel7}), that satisfy the duality condition (\ref{elmaghel5}). In this way, dynamics is encoded in the duality condition.

\section{Fourier decomposition for electromagnetic fields in vacuum} 
The method of Fourier transforms \cite{James} can be used to get solutions of Maxwell equations in vacuum (\ref{elmaghel1})-(\ref{elmaghel2}) from initial conditions for the electric and the magnetic fields. Here we review some basics for the case of electromagnetic fields defined in all the three-dimensional space $R^3$ that satisfy equations (\ref{elmaghel1})-(\ref{elmaghel2}).

Suppose that initial values of the electric and magnetic fields,
\begin{eqnarray}
{\bf E}_{0} ({\bf r} ) &=& {\bf E} ({\bf r}, t=0) , \nonumber \\
{\bf B}_{0} ({\bf r} ) &=& {\bf B} ({\bf r}, t=0) ,
\label{initialfields1} 
\end{eqnarray}
are given and defined in $R^3$. It is necessary that they satisfy the conditions
\begin{eqnarray}
\nabla \cdot {\bf E}_{0} &=& 0, \nonumber \\
\nabla \cdot {\bf B}_{0} &=& 0. \label{initialfields2}
\end{eqnarray}
These conditions assure that the electric and the magnetic fields will be divergenceless at any time, since Maxwell equations in vacuum conserve them. For example, in the case of the electric field,
\begin{equation}
\frac{\partial}{\partial t} \left( \nabla \cdot {\bf E} \right) = \nabla \cdot \left( \frac{\partial {\bf E}}{\partial t} \right) = \nabla \cdot \left( c^2 \, \nabla \times {\bf B} \right) = 0,
\label{initialfields3}
\end{equation}
where equation (\ref{elmaghel2}) has been used. A similar computation can be done for the magnetic field.

To solve Maxwell equations in vacuum with Cauchy conditions given by fields (\ref{initialfields1}) that satisfy (\ref{initialfields2}), plane wave solutions can be  proposed, such that they can be decomposed in Fourier terms,
\begin{eqnarray}
{\bf E} ({\bf r}, t) &=& \frac{1}{(2 \pi)^{3/2}} \int d^3 k \left( {\bf E}_{1} ({\bf 
k}) e^{-i ({\bf k} \cdot {\bf r} - \omega t)} +
{\bf E}_{2} ({\bf k}) e^{-i ({\bf k} \cdot {\bf r} + \omega t)} \right) , 
\nonumber \\
{\bf B} ({\bf r}, t) &=& \frac{1}{(2 \pi)^{3/2}} \int d^3 k \left( {\bf B}_{1} ({\bf 
k}) e^{-i ({\bf k} \cdot {\bf r} - \omega t)} +
{\bf B}_{2} ({\bf k}) e^{-i ({\bf k} \cdot {\bf r} + \omega t)} \right) , 
\label{elmaghel24}
\end{eqnarray}
where $\omega = c k$ and $k = \sqrt{{\bf k} \cdot {\bf k}}$. Taking $t=0$ in 
(\ref{elmaghel24}), we get
\begin{eqnarray}
{\bf E}_{0} ({\bf r}) &=& \frac{1}{(2 \pi)^{3/2}} \int d^3 k \left( {\bf E}_{1} ({\bf 
k}) + {\bf E}_{2} ({\bf k}) \right) 
e^{-i {\bf k} \cdot {\bf r}} , \nonumber \\
{\bf B}_{0} ({\bf r}) &=& \frac{1}{(2 \pi)^{3/2}} \int d^3 k \left( {\bf B}_{1} ({\bf 
k}) + {\bf B}_{2} ({\bf k}) \right) 
e^{-i {\bf k} \cdot {\bf r}} , \label{elmaghel25}
\end{eqnarray}
so that the vectors
\begin{eqnarray}
{\bf E}_{0} ({\bf k}) &=& {\bf E}_{1} ({\bf k}) + {\bf E}_{2} ({\bf k}) , \nonumber \\
{\bf B}_{0} ({\bf k}) &=& {\bf B}_{1} ({\bf k}) + {\bf B}_{2} ({\bf k}) , 
\label{elmaghel26}
\end{eqnarray}
are inverse Fourier transforms of the initial values (\ref{initialfields1}), 
\begin{eqnarray}
{\bf E}_{0} ({\bf k}) &=& \frac{1}{(2 \pi)^{3/2}} \int d^3 r \, {\bf E}_{0} ({\bf r}) 
\, e^{i {\bf k} \cdot {\bf r}} , \nonumber \\
{\bf B}_{0} ({\bf k}) &=& \frac{1}{(2 \pi)^{3/2}} \int d^3 r \, {\bf B}_{0} ({\bf r}) 
\, e^{i {\bf k} \cdot {\bf r}} ,
\label{elmaghel27}
\end{eqnarray}
and they satisfy
\begin{eqnarray}
{\bf k} \cdot {\bf E}_{0} ({\bf k}) &=& 0 , \nonumber \\ 
{\bf k} \cdot {\bf B}_{0} ({\bf k}) &=& 0,
\label{elmaghel28}
\end{eqnarray}
because of (\ref{initialfields2}).

Now, in expressions (\ref{elmaghel24}), we impose the pair of Maxwell equations $\nabla \times {\bf B} = (1/c^2) \, \partial {\bf E} / \partial t$
and $\nabla \times {\bf E} = - \partial {\bf B} / \partial t$. Taking then $t=0$ and 
using (\ref{elmaghel26}), we get the conditions
\begin{eqnarray}
{\bf k} \times {\bf E}_{0} ({\bf k}) &=& c k \left( {\bf B}_{1} ({\bf k}) - {\bf B}
_{2} ({\bf k}) \right) , \nonumber \\ 
{\bf k} \times {\bf B}_{0} ({\bf k}) &=& - \frac{k}{c} \left( {\bf E}_{1} ({\bf k}) - 
{\bf E}_{2} ({\bf k}) \right) .
\label{elmaghel29}
\end{eqnarray}
From (\ref{elmaghel29}), we obtain a solution for the vectors in $k$-space, namely
\begin{eqnarray}
{\bf E}_{1} ({\bf k}) &=& \frac{1}{2} \, {\bf E}_{0} ({\bf k}) - \frac{c}{2} \, {\bf e}
_{k} \times {\bf B}_{0} ({\bf k}) , \nonumber \\ 
{\bf E}_{2} ({\bf k}) &=& \frac{1}{2} \, {\bf E}_{0} ({\bf k}) + \frac{c}{2} \, {\bf e}
_{k} \times {\bf B}_{0} ({\bf k}) , \nonumber \\
{\bf B}_{1} ({\bf k}) &=& \frac{1}{2} \, {\bf B}_{0} ({\bf k}) + \frac{1}{2 c} \, {\bf 
e}_{k} \times {\bf E}_{0} ({\bf k}) , \nonumber
\\
{\bf B}_{2} ({\bf k}) &=& \frac{1}{2} \, {\bf B}_{0} ({\bf k}) - \frac{1}{2 c} \, {\bf 
e}_{k} \times {\bf E}_{0} ({\bf k}) ,
\label{elmaghel30}
\end{eqnarray}
where we have defined the unit vector
\begin{equation}
{\bf e}_{k} = \frac{{\bf k}}{k} . 
\label{elmaghel31}
\end{equation}
Consequently, the time-dependent fields are obtained computing first the 
Fourier transforms (\ref{elmaghel27}) of the initial values (\ref{initialfields1}), and then performing the Fourier integrals
\begin{eqnarray}
{\bf E} ({\bf r}, t) &=& \frac{1}{(2 \pi)^{3/2}} \int d^3 k \, e^{-i {\bf k} \cdot {\bf r}} \left( {\bf E}_{0} ({\bf k}) 
\cos{\omega t} - i c \, {\bf e}_{k} \times {\bf B}_{0} ({\bf k}) \sin{\omega t} \right) , \nonumber \\
{\bf B} ({\bf r}, t) &=& \frac{1}{(2 \pi)^{3/2}} \int d^3 k \, e^{-i {\bf k} \cdot {\bf r}} \left( {\bf B}_{0} ({\bf k}) 
\cos{\omega t} + \frac{i}{c} \, {\bf e}_{k} \times {\bf E}_{0} ({\bf k}) \sin{\omega t} \right) ,  \label{elmaghel32}
\end{eqnarray}
By examining equations (\ref{elmaghel26}) one can see that the complex Fourier basis components satisfy
\begin{eqnarray}
{\bf E}_{0} (- {\bf k}) &=& \bar{{\bf E}}_{0} ({\bf k}) , \nonumber \\
{\bf B}_{0} (- {\bf k}) &=& \bar{{\bf B}}_{0} ({\bf k}) , \label{elmaghel321}
\end{eqnarray}
where $\bar{{\bf E}}_{0}$ is the complex conjugate of ${\bf E}_{0}$.

The Fourier transforms can be written in other ways. We begin by taking $\cos{\omega t} = (e^{i \omega t} + e^{-i \omega t})/2$, $\sin{\omega t} = (e^{i \omega t} - e^{-i \omega t})/2i$, in equations (\ref{elmaghel32}), to get
\begin{eqnarray}
{\bf E} ({\bf r}, t) &=& \frac{1}{(2 \pi)^{3/2}} \int d^3 k \, \left[ e^{-i {\bf k} \cdot {\bf r}} e^{i \omega t} \left( \frac{1}{2} \, {\bf E}_{0} ({\bf k}) - \frac{c}{2} \, {\bf e}_{k} \times {\bf B}_{0} ({\bf k}) \right) \right. \nonumber \\
&+& \left. e^{-i {\bf k} \cdot {\bf r}} e^{- i \omega t} \left( \frac{1}{2} \, {\bf E}_{0} ({\bf k}) + \frac{c}{2} \, {\bf e}_{k} \times {\bf B}_{0} ({\bf k}) \right) \right] , \nonumber \\
{\bf B} ({\bf r}, t) &=& \frac{1}{(2 \pi)^{3/2}} \int d^3 k \, \left[ e^{-i {\bf k} \cdot {\bf r}} e^{i \omega t} \left( \frac{1}{2} \,  {\bf B}_{0} ({\bf k}) + \frac{1}{2 c} \, {\bf e}_{k} \times {\bf E}_{0} ({\bf k}) \right) \right. \nonumber \\
&+& \left. e^{-i {\bf k} \cdot {\bf r}} e^{- i \omega t} \left( \frac{1}{2} \, {\bf B}_{0} ({\bf k}) - \frac{1}{2 c} \, {\bf e}_{k} \times {\bf E}_{0} ({\bf k}) \right) \right] . \label{elmaghel322}
\end{eqnarray}
Next we perform the change ${\bf k} \mapsto - {\bf k}$ in the second terms of both fields (we can do it because the variable ${\bf k}$ is the variable of integration). Taking into account the properties (\ref{elmaghel321}), we get
\begin{eqnarray}
{\bf E} ({\bf r}, t) &=& \frac{1}{(2 \pi)^{3/2}} \int d^3 k \, \left[ e^{-i k x} \left( \frac{1}{2} \, \bar{{\bf E}}_{0} ({\bf k}) - \frac{c}{2} \, {\bf e}_{k} \times \bar{{\bf B}}_{0} ({\bf k}) \right) \right. \nonumber \\
&+& \left. e^{i k x} \left( \frac{1}{2} \, {\bf E}_{0} ({\bf k}) - \frac{c}{2} \, {\bf e}_{k} \times {\bf B}_{0} ({\bf k}) \right) \right] , \nonumber \\
{\bf B} ({\bf r}, t) &=& \frac{1}{(2 \pi)^{3/2}} \int d^3 k \, \left[ e^{-i k x} \left( \frac{1}{2} \, \bar{{\bf B}}_{0} ({\bf k}) + \frac{1}{2 c} \, {\bf e}_{k} \times \bar{{\bf E}}_{0} ({\bf k}) \right) \right. \nonumber \\
&+& \left. e^{i k x} \left( \frac{1}{2} \, {\bf B}_{0} ({\bf k}) + \frac{1}{2 c} \, {\bf e}_{k} \times {\bf E}_{0} ({\bf k}) \right) \right] , \label{elmaghel323}
\end{eqnarray}
where we have introduced the four-dimensional notation $k x = \omega t - {\bf k} \cdot {\bf r}$.

\section{Fourier transforms of vector potentials in the Coulomb gauge}
The vector potentials can also be written in this Fourier basis when we choose the Coulomb gauge for them. In the Coulomb gauge, the vector potentials are chosen so that
\begin{eqnarray}
V=0 &,& \, \, \nabla \cdot {\bf A} = 0, \nonumber \\
V^{\prime} = 0 &,& \, \, \nabla \cdot {\bf C} = 0. 
\label{coulombcondition}
\end{eqnarray}
As a consequence, they satisfy the duality equations
\begin{eqnarray}
{\bf B} &=& \nabla \times {\bf A} = \frac{1}{c^2} \, \frac{\partial {\bf C}}{\partial t} , \nonumber \\
{\bf E} &=& \nabla \times {\bf C} = - \frac{\partial {\bf A}}{\partial t} . \label{elmaghel324}
\end{eqnarray}
In this gauge, one can propose for them the following Fourier decomposition in terms of plane waves,
\begin{eqnarray}
{\bf A} ({\bf r}, t) &=& \frac{1}{(2 \pi)^{3/2}} \int d^3 k \, \left[ e^{-i k x} \, \bar{{\bf a}} ({\bf k}) + e^{i k x} \,  {\bf a} ({\bf k}) \right] , \nonumber \\
{\bf C} ({\bf r}, t) &=& \frac{c}{(2 \pi)^{3/2}} \int d^3 k \, \left[ e^{-i k x} \, \bar{{\bf c}} ({\bf k}) + e^{i k x} \,  {\bf c} ({\bf k}) \right] , \label{elmaghel325}
\end{eqnarray}
where the factor $c$ in ${\bf C}$ is taken by dimensional reasons. Taking time derivatives and using (\ref{elmaghel324}),
\begin{eqnarray}
{\bf E} &=& - \frac{\partial {\bf A}}{\partial t} =  \frac{1}{(2 \pi)^{3/2}} \int d^3 k \, \left[ e^{-i k x} \, (i c k) \, \bar{{\bf a}} ({\bf k}) - e^{i k x} \, (i c k) \, {\bf a} ({\bf k}) \right] , \nonumber \\
{\bf B} &=& \frac{1}{c^2} \, \frac{\partial {\bf C}}{\partial t} = \frac{1}{(2 \pi)^{3/2}} \int d^3 k \, \left[ - e^{-i k x} \, (i k) \, \bar{{\bf c}} ({\bf k}) + e^{i k x} \, (i k) \, {\bf c} ({\bf k}) \right] . \label{elmaghel326}
\end{eqnarray}
By comparison with equations (\ref{elmaghel323}), one gets
\begin{eqnarray}
{\bf a} ({\bf k}) &=& \frac{i}{2 c k} \left( {\bf E}_{0} ({\bf k}) - c \, {\bf e}_{k} \times {\bf B}_{0} ({\bf k}) \right), \nonumber \\ 
{\bf c} ({\bf k}) &=& \frac{-i}{2 c k} \left( c \, {\bf B}_{0} ({\bf k}) + {\bf e}_{k} \times {\bf E}_{0} ({\bf k}) \right). \label{elmaghel327}
\end{eqnarray}
Consequently, in the Coulomb gauge, the vector potentials of the electromagnetic field in vacuum can be decomposed in the Fourier basis as
\begin{eqnarray}
{\bf A} ({\bf r}, t) &=& \frac{i}{c (2 \pi)^{3/2}} \int \frac{d^3 k}{2 k} \, \left[ e^{-i k x} \left( - \bar{{\bf E}}_{0} ({\bf k}) + c \, {\bf e}_{k} \times \bar{{\bf B}}_{0} ({\bf k}) \right) \right. \nonumber \\
&+& \left. e^{i k x} \,  \left( {\bf E}_{0} ({\bf k}) - c \, {\bf e}_{k} \times {\bf B}_{0} ({\bf k}) \right) \right] , \nonumber \\
{\bf C} ({\bf r}, t) &=& \frac{i}{(2 \pi)^{3/2}} \int \frac{d^3 k}{2 k} \, \left[ e^{-i k x} \, \left( c \, \bar{{\bf B}}_{0} ({\bf k}) + {\bf e}_{k} \times \bar{{\bf E}}_{0} ({\bf k}) \right) \right. \nonumber \\
&+& \left. e^{i k x} \,  \left( - c \, {\bf B}_{0} ({\bf k}) - {\bf e}_{k} \times {\bf E}_{0} ({\bf k}) \right) \right] . \label{elmaghel328}
\end{eqnarray} 

\section{Helicity basis}
The helicity Fourier components appear when the vector potentials ${\bf A}$ and ${\bf C}$, in the Coulomb gauge, are written as a combination of circularly polarized waves \cite{Ynd}, as
\begin{eqnarray}
{\bf A} ({\bf r}, t) &=& \frac{\sqrt{\hbar c \mu_{0}}}{(2 \pi )^{3/2}} \int \frac{d^{3} k}{\sqrt{2 k}} \left[ e^{-i k x} \left( a_{R} ({\bf k})
{\bf e}_{R} ({\bf k}) + a_{L} ({\bf k}) {\bf e}_{L} ({\bf k}) \right)  + C.C. \right] , \nonumber \\
{\bf C} ({\bf r}, t) &=& \frac{c \sqrt{\hbar c \mu_{0}}}{(2 \pi )^{3/2}} \int \frac{d^{3} k}{\sqrt{2 k}} \left[ i \, e^{-i k x} \left( a_{R} 
({\bf k}) {\bf e}_{R} ({\bf k}) - a_{L} ({\bf k}) {\bf e}_{L} ({\bf k}) \right) + C.C. \right] .
\label{elmaghel34}
\end{eqnarray}
where $\hbar$ is the Planck constant, $\mu_{0}$ is the vacuum magnetic permeability and $C.C.$ means complex conjugate. The Fourier components in the helicity basis are given by the complex unit vectors ${\bf e}_{R} ({\bf k})$, ${\bf e}_{L} ({\bf k})$, ${\bf e}_{k} = {\bf k}/k$, and the helicity components $a_{R} ({\bf k})$, $a_{L} ({\bf k})$ that, in the quantum theory, are interpreted as annihilation operators of photon states with right- and left-handed polarization, respectively ($\bar{a}_{R} ({\bf k})$, $\bar{a}_{L} ({\bf k})$ are, in quantum theory, creation operators of such states) \cite{Ynd}.

The unit vectors in the helicity basis are taken to satisfy the relations,
\begin{eqnarray}
\bar{{\bf e}}_{R} = {\bf e}_{L} , \, {\bf e}_{R} (- {\bf k}) = - {\bf e}_{L} ({\bf k}) , \, {\bf e}_{L} 
(- {\bf k}) = - {\bf e}_{R} ({\bf k}) , \nonumber \\
{\bf e}_{k} \cdot {\bf e}_{R} = {\bf e}_{k} \cdot {\bf e}_{L} = 0 , \, {\bf e}_{R}
\cdot {\bf e}_{R} = {\bf e}_{L} \cdot {\bf e}_{L} = 0 , \, {\bf e}_{R} \cdot {\bf e}_{L} = 1, \nonumber \\
{\bf e}_{k} \times {\bf e}_{k} = {\bf e}_{R} \times {\bf e}_{R} = {\bf e}_{L}
\times {\bf e}_{L} = 0 , \nonumber \\
{\bf e}_{k} \times {\bf e}_{R} = -i \, {\bf e}_{R} , \, {\bf e}_{k} \times {\bf e}_{L} = i 
\, {\bf e}_{L} , \, {\bf e}_{R}\times {\bf e}_{L}= -i \, {\bf e}_{k}. \label{elmaghel35}
\end{eqnarray}
These relations can be achieved by taking real vectors ${\bf e}_{1}$, ${\bf e}_{2}$ in $k$-space such that
\begin{eqnarray}
{\bf e}_{1} \cdot {\bf e}_{1} = {\bf e}_{2} \cdot {\bf e}_{2} = \frac{1}{2} , \, {\bf e}_{1} \cdot {\bf e}_{2} = {\bf e}_{1} \cdot {\bf e}_{k} = {\bf e}_{2} \cdot {\bf e}_{k} = 0 , \nonumber \\
{\bf e}_{1} \times {\bf e}_{2} = \frac{1}{2} \, {\bf e}_{k} , \, {\bf e}_{k} \times {\bf e}_{1} = {\bf e}_{2} , \, {\bf e}_{2} \times {\bf e}_{k} = {\bf e}_{1} , \nonumber \\
{\bf e}_{1} (- {\bf k}) = - {\bf e}_{1} ({\bf k}) , \, {\bf e}_{2} (- {\bf k}) = {\bf e}_{2} ({\bf k}) ,
\label{elmaghel35b}
\end{eqnarray}
and constructing the complex unit vectors
\begin{eqnarray}
{\bf e}_{R} &=& {\bf e}_{1} + i \, {\bf e}_{2} , \, \nonumber \\
{\bf e}_{L} &=& {\bf e}_{1} - i \, {\bf e}_{2} .
\label{elmaghel35c}
\end{eqnarray}

The relation between the helicity basis and the planar Fourier basis can be understood by comparing equations (\ref{elmaghel328}) and (\ref{elmaghel34}). It is found that:
\begin{eqnarray}
a_{R} {\bf e}_{R} + a_{L} {\bf e}_{L} &=& \frac{1}{c \sqrt{\hbar c \mu_{0}} \sqrt{2k}} \, \left( - i \, \bar{{\bf E}}_{0} ({\bf k}) + i c \, {\bf e}_{k} \times \bar{{\bf B}}_{0} ({\bf k}) \right) , \nonumber \\
a_{R} {\bf e}_{R} - a_{L} {\bf e}_{L} &=& \frac{1}{c \sqrt{\hbar c \mu_{0}} \sqrt{2k}} \left( c \, \bar{{\bf
B}}_{0} ({\bf k}) + {\bf e}_{k} \times \bar{{\bf E}}_{0} ({\bf k}) \right) . \label{elmaghel39}
\end{eqnarray}
From these expressions,
\begin{eqnarray}
a_{R} {\bf e}_{R} &=& \frac{1}{2 \sqrt{\hbar c \mu_{0}} \sqrt{2k}} \left[ \left( \bar{{\bf B}}_{0} ({\bf k})- \frac{i}{c} \, \bar{{\bf E}}_{0} ({\bf k}) \right) + i  {\bf e}_{k} \times \left( \bar{{\bf B}}_{0} ({\bf k})- \frac{i}{c} \, \bar{{\bf E}}_{0} ({\bf k}) \right) \right] , \nonumber \\
a_{L} {\bf e}_{L} &=& \frac{-1}{2 \sqrt{\hbar c \mu_{0}} \sqrt{2k}} \left[ \left( \bar{{\bf B}}_{0} ({\bf k}) + \frac{i}{c} \, \bar{{\bf E}}_{0} ({\bf k}) \right) - i  {\bf e}_{k} \times \left( \bar{{\bf B}}_{0} ({\bf k})+ \frac{i}{c} \, \bar{{\bf E}}_{0} ({\bf k}) \right) \right] . \label{elmaghel391}
\end{eqnarray}
Consequently, in the helicity basis fields in vacuum are written as superposition of circularly polarized waves. The electric and magnetic fields of an electromagnetic field in vacuum, and the vector potentials in the Coulomb gauge, can be expressed in this basis as (we do not write the dependence on ${\bf k}$ in the following expressions for clarity):
\begin{eqnarray}
{\bf E} ({\bf r}, t) &=& \frac{c \sqrt{\hbar c \mu_{0}}}{(2 \pi )^{3/2}} \int d^{3} k \, \sqrt{\frac{k}{2}} \, \left[ i \, e^{-i k x} \left( a_{R} {\bf e}_{R} + a_{L} {\bf e}_{L} \right)  - i \, e^{i k x} \left( \bar{a}_{R} {\bf e}_{L} + \bar{a}_{L} {\bf e}_{R} \right) \right] , \nonumber \\
{\bf B} ({\bf r}, t) &=& \frac{\sqrt{\hbar c \mu_{0}}}{(2 \pi )^{3/2}} \int d^{3} k \, \sqrt{\frac{k}{2}} \, \left[ e^{-i k x} \left( a_{R} {\bf e}_{R} - a_{L} {\bf e}_{L} \right) + e^{i k x} \left( \bar{a}_{R} {\bf e}_{L} - \bar{a}_{L} {\bf e}_{R} \right) \right] , \nonumber \\
{\bf A} ({\bf r}, t) &=& \frac{\sqrt{\hbar c \mu_{0}}}{(2 \pi )^{3/2}} \int d^{3} k \, \frac{1}{\sqrt{2 k}} \, \left[ e^{-i k x} \left( a_{R} {\bf e}_{R} + a_{L} {\bf e}_{L} \right) + e^{i k x} \left( \bar{a}_{R} {\bf e}_{L} + \bar{a}_{L} {\bf e}_{R} \right) \right] , \nonumber \\
{\bf C} ({\bf r}, t) &=& \frac{c \sqrt{\hbar c \mu_{0}}}{(2 \pi )^{3/2}} \int d^{3} k \, \frac{1}{\sqrt{2 k}} \, \left[ i \, e^{-i k x} \left( a_{R} {\bf e}_{R} - a_{L} {\bf e}_{L} \right) - i \, e^{i k x} \left( \bar{a}_{R} {\bf e}_{L} - \bar{a}_{L} {\bf e}_{R} \right) \right] . \nonumber \\
\label{elmaghel392}
\end{eqnarray}
The unit vectors in this basis satisfy equations (\ref{elmaghel35}) and the basis elements are related to the planar Fourier basis through equations (\ref{elmaghel39}) or (\ref{elmaghel391}).

Helicity basis can be used in relation to the helicity of a vector field \cite{Moffatt,Berger,Ricca,Berger1999,Dennis05,Ricca2011}, that is a useful quantity in the study of topological configurations of electric and magnetic lines. In the case of electromagnetism in vacuum, the magnetic helicity can be defined as the integral
\begin{equation}
h_{m}= \frac{1}{2 c \mu_{0}} \int d^3 r \, {\bf A} \cdot {\bf B} ,
\label{elmaghel10}
\end{equation}
and the electric helicity \cite{Afana,True96,Ran01,Bliokh2013} is 
\begin{equation}
h_{e} = \frac{\varepsilon_{0}}{2c} \int d^3 r \, {\bf C} \cdot {\bf E} = \frac{1}{2c^3 \mu_{0}} \int d^3 r \, {\bf C} \cdot {\bf E},
\label{elmaghel11}
\end{equation}
where $\varepsilon_{0} = 1/(c^2 \mu_{0} )$ is the vacuum electric permittivity. It can be proved \cite{Arr12,Arr18} that (i) if the integral of ${\bf E} \cdot {\bf B}$ is zero, both the magnetic and the electric helicities are constant during the evolution
of the electromagnetic field, (ii) if the integral of ${\bf E} \cdot {\bf B}$ is not zero, the helicities are not constant but they satisfy $d h_{m}/dt = - d h_{e}/dt$, so we get an interchange of helicities between the magnetic and electric parts of the field, and (iii) for every value of the integral of ${\bf E} \cdot {\bf B}$, the electromagnetic helicity $h = h_{m} + h_{e}$ is a conserved quantity.

Using the helicity basis defined in (\ref{elmaghel34}), the electromagnetic helicity $h$ of an electromagnetic field in vacuum can be written as
\begin{equation}
h = h_{m} + h_{e} = \hbar \int d^3 k \, \left( \bar{a}_{R}({\bf k}) a_{R} ({\bf k}) - \bar{a}_{L} ({\bf k}) a_{L} ({\bf k}) \right).
\label{elmaghel45}
\end{equation}
From the usual expressions for the number of right- and left-handed photons in Quantum Mechanics,
\begin{eqnarray}
N_{R} &=& \int d^3 k \, \bar{a}_{R}({\bf k}) a_{R} ({\bf k}) , \nonumber \\
N_{L} &=& \int d^3 k \, \bar{a}_{L}({\bf k}) a_{L} ({\bf k}) ,
\label{elmaghel45b}
\end{eqnarray}
we can write
\begin{equation}
h = \hbar \left( N_{R}-N_{L} \right) .
\label{elmaghel46}
\end{equation}
Consequently, the electromagnetic helicity is the classical limit of the difference between the numbers of right-handed and left-handed photons \cite{Afana,True96,Ran97}.

\section{Fourier method in the case of a set of knotted electromagnetic fields}
There is a set of so-called non-null torus electromagnetic knots \cite{Arr15,Arr17} that are exact solutions of Maxwell equations in vacuum with the property that, at a given initial time $t = 0$, all pairs of lines of the field ${\bf B} ({\bf r},0)$ are linked torus knots, and that the linking number is the same for all the pairs. Similarly, all pairs of lines of the field ${\bf E} ({\bf r},0)$ are linked torus knots and the linking number of all pairs of lines is the same.

Take four positive integers $(n, m, l, s)$. Then it is possible to give an initial magnetic field such that all its magnetic lines are $(n,m)$ torus knots. This initial magnetic field is \cite{Arr15}
\begin{equation}
{\bf B}_{0} ({\bf r}) = \frac{8 \sqrt{a}}{\pi L_{0}^2 (1 + R^2)^3} \left( m Y - n X Z , - m X - n Y Z , n \frac{X^2 + Y^2 - Z^2 -1}{2} \right) , 
\label{fourierknots1}
\end{equation}
where $a$ is some constant proportional to the product $\hbar c \mu_{0}$ to give the magnetic field its correct dimensions in SI units, $L_{0}$ is a constant with dimensions of length, and we define the dimensionless coordinates $(X, Y, Z)$, related to the physical ones $(x, y, z)$ in the SI of units by
\begin{equation}
(X, Y, Z) = \frac{(x, y, z)}{L_{0}} ,
\label{fourierknots1b}
\end{equation}
so that
\begin{equation}
R^2 = X^2 + Y^2 +Z^2 = \frac{x^2 + y^2 +z^2}{L_{0}^2} = \frac{r^2}{L_{0}^2} .
\label{fourierknots1c}
\end{equation}
The linking number of every two of magnetic lines at $t=0$ is equal to $nm$. Similarly, the initial electric field
\begin{equation}
{\bf E}_{0} ({\bf r}) = \frac{8 c \sqrt{a}}{\pi L_{0}^2 (1 + R^2)^3} \left( l \frac{X^2 - Y^2 - Z^2 +1}{2} , l X Y - s Z , l X Z + s Y , \right) 
\label{fourierknots2}
\end{equation}
is such that all its electric lines are $(l,s)$ torus knots and any pair of electric lines has a linking number equal to $l s$. These topological properties only happen for $t=0$. This is due to the fact that topology changes during time evolution if one of the integers $(n, m, l, s)$ is different from any of the others (see \cite{Arr15}).

Let us compute the magnetic and electric fields from the Cauchy conditions (\ref{fourierknots1}) and (\ref{fourierknots2}), that satisfy the required conditions $\nabla \cdot {\bf B}_{0} = \nabla \cdot {\bf E}_{0} = 0$. We will use the dimensionless time $T = c t / L_{0}$. Before the computations, we can use (\ref{elmaghel32}) to build the complex vector field
\begin{eqnarray}
{\bf B} ({\bf r}, t) + i \, \frac{{\bf E} ({\bf r}, t)}{c} &=& \frac{1}{(2 \pi)^{3/2}} \int d^3 k \, e^{-i {\bf k} \cdot {\bf r}} \, \left[ \left( {\bf B}_{0} ({\bf k}) + i \, \frac{{\bf E}_{0} ({\bf k})}{c} \right) \cos{\omega t} \right. \nonumber \\ 
&+& \left. {\bf e}_{k} \times \left( {\bf B}_{0} ({\bf k}) + i \, \frac{{\bf E}_{0} ({\bf k})}{c} \right) \sin{\omega t} \right] , 
\label{fourierknots3}
\end{eqnarray}
to show that we need first to compute the complex vector field ${\bf B}_{0} ({\bf k}) + i/c \, {\bf E}_{0} ({\bf k})$. Using (\ref{elmaghel27}), we get
\begin{equation}
{\bf B}_{0} ({\bf k}) + i \, \frac{{\bf E}_{0} ({\bf k})}{c} = \frac{1}{(2 \pi)^{3/2}} \int d^3 r \, e^{i {\bf k} \cdot {\bf r}} \, \left( {\bf B}_{0} ({\bf r}) + i \, \frac{{\bf E}_{0} ({\bf r})}{c} \right) ,
\label{fourierknots4}
\end{equation}
where ${\bf B}_{0} ({\bf r})$ and ${\bf E}_{0} ({\bf r})$ are given by (\ref{fourierknots1}) and (\ref{fourierknots2}) respectively. We are going to use dimensionless coordinates in $k$-space, defined by
\begin{equation}
(K_{x} , K_{y} , K_{z}) = L_{0} \, (k_{x} , k_{y}, k_{z}) , \, \, K = L_{0} k = \frac{L_{0} \omega}{c} .
\label{fourierknots5}
\end{equation}
The integrals in (\ref{fourierknots4}) can be done to give
\begin{eqnarray}
{\bf B}_{0} ({\bf k}) + i \, \frac{{\bf E}_{0} ({\bf k})}{c} &=& \frac{L_{0} \sqrt{a} \, e^{-K}}{\sqrt{2 \pi}} \, \left[ \frac{n}{K} \left( K_{x} K_{z} , K_{y} K_{z} , - K_{x}^2 - K_{y}^2 \right) \right. \nonumber \\ 
&+& \left. i \, m \left( K_{y} , - K_{x} , 0 \right) \right. \nonumber \\
&+& \left. i \, \frac{l}{K} \left( K_{y}^2 + K_{z}^2 , - K_{x} K_{y} , - K_{x} K_{z} \right) \right. \nonumber \\
&+& \left. s \left( 0 , K_{z} , - K_{y} \right) \right] .
\label{fourierknots6}
\end{eqnarray}
so the Fourier transform of the initial magnetic field ${\bf B}_{0} ({\bf r})$ is
\begin{eqnarray}
{\bf B}_{0} ({\bf k}) &=& \frac{L_{0} \sqrt{a} \, e^{-K}}{\sqrt{2 \pi}} \, \left[ \frac{n}{K} \left( K_{x} K_{z} , K_{y} K_{z} , - K_{x}^2 - K_{y}^2 \right) \right. \nonumber \\ 
&+& \left. i \, m \left( K_{y} , - K_{x} , 0 \right) \right] ,
\label{fourierknots6b}
\end{eqnarray}
and the Fourier transform of the initial magnetic field ${\bf E}_{0} ({\bf r})$ is
\begin{eqnarray}
{\bf E}_{0} ({\bf k}) &=& \frac{L_{0} c \sqrt{a} \, e^{-K}}{\sqrt{2 \pi}} \, \left[ \frac{l}{K} \left( K_{y}^2 + K_{z}^2 , - K_{x} K_{y} , - K_{x} K_{z} \right) \right. \nonumber \\
&+& \left. i\, s \left( 0 , - K_{z} , K_{y} \right) \right] .
\label{fourierknots6c}
\end{eqnarray}
Note that the Fourier transforms (\ref{fourierknots6b}) and (\ref{fourierknots6c}) satisfy the conditions ${\bf B}_{0} (- {\bf k}) = \bar{{\bf B}}_{0} ({\bf k})$ and ${\bf E}_{0} (- {\bf k}) = \bar{{\bf E}}_{0} ({\bf k})$, as was established in (\ref{elmaghel321}). According to (\ref{fourierknots3}), we also need to compute, using (\ref{fourierknots6}), the complex vector 
\begin{eqnarray}
{\bf e}_{k} \times \left( {\bf B}_{0} ({\bf k}) + i \, \frac{{\bf E}_{0} ({\bf k})}{c} \right) &=& \frac{L_{0} \sqrt{a} \, e^{-K}}{\sqrt{2 \pi}} \, \left[ n \left( - K_{y} , K_{x} , 0 \right) \right. \nonumber \\ 
&+& \left. i \, \frac{m}{K} \left( K_{x} K_{z} , K_{y} K_{z} , - K_{x}^2 - K_{y}^2 \right) \right. \nonumber \\
&+& \left. i \, l \left( 0 , K_{z} , - K_{y} \right) \right. \nonumber \\
&+& \left. \frac{s}{K} \left( - K_{y}^2 - K_{z}^2 , K_{x} K_{y} , K_{x} K_{z} \right) \right] .
\label{fourierknots7}
\end{eqnarray}
The next goal is to compute the integral (\ref{fourierknots3}) that, in terms of dimensionless coordinates in $x$-space and $k$-space, can be written as
\begin{eqnarray}
{\bf B} ({\bf r}, t) + i \, \frac{{\bf E} ({\bf r}, t)}{c} &=& \frac{1}{(2 \pi)^{3/2} L_{0}^3} \int d^3 K \, e^{-i {\bf K} \cdot {\bf R}} \, \left[ \left( {\bf B}_{0} ({\bf K}) + i \, \frac{{\bf E}_{0} ({\bf K})}{c} \right) \cos{K T} \right. \nonumber \\ 
&+& \left. \frac{{\bf K}}{K} \times \left( {\bf B}_{0} ({\bf K}) + i \, \frac{{\bf E}_{0} ({\bf K})}{c} \right) \sin{K T} \right] .
\label{fourierknots8}
\end{eqnarray}
It is interesting to compute first the following integrals,
\begin{eqnarray}
I_{1c} &=& \frac{\sqrt{a}}{4 \pi^2 L_{0}^2} \int d^3 K \, e^{-i {\bf K} \cdot {\bf R}} \, \cos{KT} \, \frac{e^{-K}}{K} = \frac{\sqrt{a}}{2 \pi L_{0}^2} \, \frac{A}{A^2 + T^2}  , \nonumber \\
I_{2c} &=& \frac{\sqrt{a}}{4 \pi^2 L_{0}^2} \int d^3 K \, e^{-i {\bf K} \cdot {\bf R}} \, \cos{KT} = \frac{\sqrt{a}}{2 \pi L_{0}^2} \, \frac{A^2 - T^2 + 2 A T^2}{(A^2 + T^2)^2}  , \nonumber \\
I_{1s} &=& \frac{\sqrt{a}}{4 \pi^2 L_{0}^2} \int d^3 K \, e^{-i {\bf K} \cdot {\bf R}} \, \sin{KT} \, \frac{e^{-K}}{K} = \frac{\sqrt{a}}{2 \pi L_{0}^2} \, \frac{T}{A^2 + T^2}  , \nonumber \\
I_{2s} &=& \frac{\sqrt{a}}{4 \pi^2 L_{0}^2} \int d^3 K \, e^{-i {\bf K} \cdot {\bf R}} \, \sin{KT} = \frac{\sqrt{a}}{2 \pi L_{0}^2} \, \frac{T^3 + 2 A T - A^2 T}{(A^2 + T^2)^2}  ,
\label{fourierknots9}
\end{eqnarray}
in which
\begin{equation}
A = \frac{R^2 - T^2 +1}{2} .
\label{fourierknots10}
\end{equation}
Now we make one useful observation. If, in the integral (\ref{fourierknots8}), we need to compute a term containing $K_{x}$ times one of the integrands of the expressions (\ref{fourierknots9}), we can do it in this way,
\begin{equation}
\frac{\sqrt{a}}{4 \pi^2 L_{0}^2} \int d^3 K \, e^{-i {\bf K} \cdot {\bf R}} \, \cos{KT} \, \frac{e^{-K}}{K} \, K_{x} = i \, \frac{\partial I_{1c}}{\partial X} = i X \, \frac{d I_{1c}}{dA} ,
\label{fourierknots11}
\end{equation}
and the same happens with $I_{2c}$, $I_{1s}$ and $I_{2s}$. If we need to compute a term containing $K_{x}^2$, we can make 
\begin{equation}
\frac{\sqrt{a}}{4 \pi^2 L_{0}^2} \int d^3 K \, e^{-i {\bf K} \cdot {\bf R}} \, \cos{KT} \, \frac{e^{-K}}{K} \, K_{x}^2 = - \frac{d I_{1c}}{dA} - X^2 \, \frac{d^2 I_{1c}}{dA^2} ,
\label{fourierknots12}
\end{equation}
and, if we need to compute a term containing $K_{x} K_{y}$, 
\begin{equation}
\frac{\sqrt{a}}{4 \pi^2 L_{0}^2} \int d^3 K \, e^{-i {\bf K} \cdot {\bf R}} \, \cos{KT} \, \frac{e^{-K}}{K} \, K_{x} K_{y} = - X Y \, \frac{d^2 I_{1c}}{dA^2} .
\label{fourierknots13}
\end{equation}
Defining the quantities
\begin{eqnarray}
P &=& T (T^2 - 3 A^2 ) , \nonumber \\
Q &=& A (A^2 - 3 T^2 ) ,
\label{fourierknots14}
\end{eqnarray}
we can finally compute (\ref{fourierknots8}), obtaining that the set of non-null torus electromagnetic knots can be written as  
\begin{eqnarray}
{\bf B} ({\bf r}, t) &=& \frac{\sqrt{a}}{\pi L_{0}^2} \, \frac{Q \, {\bf H}_{1} + P \, {\bf H}_{2}}{(A^2 + T^2 )^3} \label{knot10} \\
{\bf E} ({\bf r}, t) &=& \frac{\sqrt{a} c}{\pi L_{0}^2} \, \frac{Q \, {\bf H}_{4} - P \, {\bf H}_{3}}{(A^2 + T^2 )^3} \label{knot11}
\end{eqnarray}
where
\begin{eqnarray}
{\bf H}_{1} = \left( -n XZ + m Y + s T , -n YZ -m X -l TZ , n \frac{-1 - Z^2 + X^2 + Y^2 +T^2}{2} + l TY \right) . \nonumber \\
{\bf H}_{2} = \left( s \frac{1+X^2 -Y^2-Z^2-T^2}{2} -m TY , s XY - l Z + m TX , s XZ + l Y + n T \right) . \nonumber \\
{\bf H}_{3} = \left( -m XZ + n Y + l T , -m YZ -n X -s TZ ,  m \frac{-1 - Z^2 + X^2 + Y^2 +T^2}{2} + s TY \right) . \nonumber \\
{\bf H}_{4} = \left( l \frac{1+X^2 -Y^2-Z^2-T^2}{2} - n TY , l XY - s Z + n TX , l XZ + s Y + m T \right) . \label{knot16}
\end{eqnarray}

\section{Helicity basis for the set of electromagnetic knots}
Let us consider the helicity basis for the electromagnetic fields we have computed, since it is interesting to understand some physical quantities of the electromagnetic field such as energy, helicity and linear and angular momentum. Using expressions (\ref{fourierknots6b}) and (\ref{fourierknots6c}), we get
\begin{eqnarray}
\bar{{\bf B}}_{0} ({\bf k}) &=& \frac{L_{0} \sqrt{a} \, e^{-K}}{\sqrt{2 \pi}} \, \left[ \frac{n}{K} \left( K_{x} K_{z} , K_{y} K_{z} , - K_{x}^2 - K_{y}^2 \right) \right. \nonumber \\ 
&-& \left. i \, m \left( K_{y} , - K_{x} , 0 \right) \right] , \nonumber \\
\bar{{\bf E}}_{0} ({\bf k}) &=& \frac{L_{0} c \sqrt{a} \, e^{-K}}{\sqrt{2 \pi}} \, \left[ \frac{l}{K} \left( K_{y}^2 + K_{z}^2 , - K_{x} K_{y} , - K_{x} K_{z} \right) \right. \nonumber \\
&-& \left. i\, s \left( 0 , - K_{z} , K_{y} \right) \right] ,
\label{knots20c}
\end{eqnarray}
so that
\begin{eqnarray}
\bar{{\bf B}}_{0} ({\bf k}) - \frac{i}{c} \, \bar{{\bf E}}_{0} ({\bf k})&=& \frac{L_{0} \sqrt{a} \, e^{-K}}{\sqrt{2 \pi}} \, \left[ \frac{n}{K} \left( K_{x} K_{z} , K_{y} K_{z} , - K_{x}^2 - K_{y}^2 \right) \right. \nonumber \\ 
&+& \left. s \left( 0 , K_{z} , - K_{y} \right) \right. \nonumber \\
&-& \left. i \, m \left( K_{y} , - K_{x} , 0 \right) \right. , \nonumber \\
&-& \left. i \, \frac{l}{K} \left( K_{y}^2 + K_{z}^2 , - K_{x} K_{y} , - K_{x} K_{z} \right) \right] ,
\label{knots20d}
\end{eqnarray}
and
\begin{eqnarray}
{\bf e}_{k} \times \left( \bar{{\bf B}}_{0} ({\bf k}) - \frac{i}{c} \, \bar{{\bf E}}_{0} ({\bf k}) \right) &=& \frac{L_{0} \sqrt{a} \, e^{-K}}{\sqrt{2 \pi}} \, \left[ - n \left( K_{y} , - K_{x} , 0 \right) \right. \nonumber \\ 
&-& \left. \frac{s}{K} \left( K_{y}^2 + K_{z}^2 , - K_{x} K_{y}, - K_{x} K_{z} \right) \right] , \nonumber \\
&-& \left. i \, \frac{m}{K} \left( K_{x} K_{z} , K_{y} K_{z} , - K_{x}^2 - K_{y}^2 \right) \right. \nonumber \\
&-& \left. i \, l \left( 0 , K_{z} , - K_{y} \right) \right] .
\label{knots20e}
\end{eqnarray}
According to (\ref{elmaghel391}),
\begin{eqnarray}
a_{R} {\bf e}_{R} &=& \sqrt{\frac{a}{\hbar c \mu_{0}}} \, \frac{L_{0}^{3/2}}{4 \sqrt{\pi}} \, \frac{e^{-K}}{\sqrt{K}} \, \times \nonumber \\
& &\left[ \frac{n+m}{K} \left( K_{x} K_{z}, K_{y} K_{z}, - K_{x}^2 -K_{y}^2 \right) + (l + s) \left( 0, K_{z}, - K_{y} \right) \right] \nonumber \\
&-& i \, \left[ \frac{l+s}{K} \left( K_{y}^2 + K_{z}^2, - K_{x}K_{y}, - K_{x}K_{z} \right) + (n + m) \left( K_{y}, - K_{x},0 \right) \right] . \label{knot21}
\end{eqnarray}
Similarly,
\begin{eqnarray}
\bar{{\bf B}}_{0} ({\bf k}) + \frac{i}{c} \, \bar{{\bf E}}_{0} ({\bf k})&=& \frac{L_{0} \sqrt{a} \, e^{-K}}{\sqrt{2 \pi}} \, \left[ \frac{n}{K} \left( K_{x} K_{z} , K_{y} K_{z} , - K_{x}^2 - K_{y}^2 \right) \right. \nonumber \\ 
&-& \left. s \left( 0 , K_{z} , - K_{y} \right) \right. \nonumber \\
&-& \left. i \, m \left( K_{y} , - K_{x} , 0 \right) \right. , \nonumber \\
&+& \left. i \, \frac{l}{K} \left( K_{y}^2 + K_{z}^2 , - K_{x} K_{y} , - K_{x} K_{z} \right) \right] ,
\label{knots22d}
\end{eqnarray}
and
\begin{eqnarray}
{\bf e}_{k} \times \left( \bar{{\bf B}}_{0} ({\bf k}) + \frac{i}{c} \, \bar{{\bf E}}_{0} ({\bf k}) \right) &=& \frac{L_{0} \sqrt{a} \, e^{-K}}{\sqrt{2 \pi}} \, \left[ - n \left( K_{y} , - K_{x} , 0 \right) \right. \nonumber \\ 
&+& \left. \frac{s}{K} \left( K_{y}^2 + K_{z}^2 , - K_{x} K_{y}, - K_{x} K_{z} \right) \right] , \nonumber \\
&-& \left. i \, \frac{m}{K} \left( K_{x} K_{z} , K_{y} K_{z} , - K_{x}^2 - K_{y}^2 \right) \right. \nonumber \\
&+& \left. i \, l \left( 0 , K_{z} , - K_{y} \right) \right] ,
\label{knots22e}
\end{eqnarray}
from which
\begin{eqnarray}
a_{L} {\bf e}_{L} &=& \sqrt{\frac{a}{\hbar c \mu_{0}}} \, \frac{L_{0}^{3/2}}{4 \sqrt{\pi}} \, \frac{e^{-K}}{\sqrt{K}} \, \times \nonumber \\
& &\left[ \frac{m-n}{K} \left( K_{x} K_{z}, K_{y} K_{z}, - K_{x}^2 -K_{y}^2 \right) + (s - l) \left( 0, K_{z}, - K_{y} \right) \right] \nonumber \\
&-& i \, \left[ \frac{l-s}{K} \left( K_{y}^2 + K_{z}^2, - K_{x}K_{y}, - K_{x}K_{z} \right) + (n-m) \left( K_{y}, - K_{x},0 \right) \right] . \label{knot23}
\end{eqnarray}

One interesting application of these results appears by multiplying $a_{R} {\bf e}_{R}$ and its complex conjugate, and the same for $a_{L} {\bf e}_{L}$. Using (\ref{elmaghel35}),
\begin{eqnarray}
(\bar{a}_{R} \bar{\bf e}_{R}) \cdot (a_{R} {\bf e}_{R}) &=& \bar{a}_{R} a_{R} \left( {\bf e}_{L} \cdot {\bf e}_{R} \right) = \bar{a}_{R} a_{R} , \nonumber \\
(\bar{a}_{L} \bar{\bf e}_{L}) \cdot (a_{L} {\bf e}_{L}) &=& \bar{a}_{L} a_{L} \left( {\bf e}_{R} \cdot {\bf e}_{L} \right) = \bar{a}_{L} a_{L} .
\label{knot24}
\end{eqnarray}
In our case,
\begin{eqnarray}
\bar{a}_{R} a_{R} &=& \frac{a}{\hbar c \mu_{0}} \frac{L_{0}^3}{8 \pi} \frac{e^{-2 K}}{K} \left( (n + m)^2 (K_{x}^2 + K_{y}^2) + (l + s)^2 (K_{y}^2 + K_{z}^2) \right. \nonumber \\
&+& \left. 4 (n + m) (l + s) K K_{y} \right), \nonumber \\
\bar{a}_{L} a_{L} &=& \frac{a}{\hbar c \mu_{0}} \frac{L_{0}^3}{8 \pi} \frac{e^{-2 K}}{K} \left( (n - m)^2 (K_{x}^2 + K_{y}^2) + (l - s)^2 (K_{y}^2 + K_{z}^2) \right. \nonumber \\
&+& \left. 4 (n - m) (l - s) K K_{y} \right) .
\label{knot25}
\end{eqnarray}
From (\ref{elmaghel45b}), the classical expressions corresponding to the quantum mechanical number of right- and left-handed photons for our set of electromagnetic knots are
\begin{eqnarray}
N_{R} &=& \int d^3 k \, \bar{a}_{R} a_{R} = \frac{1}{L_{0}^3} \int d^3 K \, \bar{a}_{R} a_{R} = \frac{a}{\hbar c \mu_{0}} \frac{(n+m)^2 + (l +s)^2}{8}, \nonumber \\
N_{L} &=& \int d^3 k \, \bar{a}_{L} a_{L} = \frac{1}{L_{0}^3} \int d^3 K \, \bar{a}_{L} a_{L} = \frac{a}{\hbar c \mu_{0}} \frac{(n-m)^2 + (l -s)^2}{8},
\label{knot26}
\end{eqnarray}
where we have to remember that the constant $a$ is proportional to the product $\hbar c \mu_{0}$ because of dimensional reasons. In the particular case of the Hopfion, for which $n = m = l = s = 1$, we have
\begin{eqnarray}
N_{R} (\mbox{Hopfion}) &=& \frac{a}{\hbar c \mu_{0}} , \nonumber \\
N_{L} (\mbox{Hopfion}) &=& 0 .
\label{knot27}
\end{eqnarray}
If, moreover, we set the arbitrary constant $a$ to have the value $a = \hbar c \mu_{0}$, for the set of electromagnetic knots we are studying we get
\begin{eqnarray}
N_{R} &=& \int d^3 k \, \bar{a}_{R} a_{R} = \frac{1}{L_{0}^3} \int d^3 K \, \bar{a}_{R} a_{R} = \frac{(n+m)^2 + (l +s)^2}{8}, \nonumber \\
N_{L} &=& \int d^3 k \, \bar{a}_{L} a_{L} = \frac{1}{L_{0}^3} \int d^3 K \, \bar{a}_{L} a_{L} = \frac{(n-m)^2 + (l -s)^2}{8},
\label{knot28}
\end{eqnarray}
and for the Hopfion,
\begin{eqnarray}
N_{R} (\mbox{Hopfion}) &=& 1 , \nonumber \\
N_{L} (\mbox{Hopfion}) &=& 0 ,
\label{knot29}
\end{eqnarray}
so the classical value corresponding to the number of right-handed photons in the Hopfion is 1, and the classical value corresponding to the number of left-handed photons is 0.

\section{Conclusions}
Solutions to Maxwell equations in vacuum can be obtained by taking an initial electromagnetic field configuration, and by Fourier transform, propagating it in time. This construction is used in the generation of non-null toroidal electromagnetic fields \cite{Arr15}. Here we have presented a detailed calculation suitable for students with a basic mathematical background.

The helicity is a property which characterizes some non-trivial topological behaviour of the field and can be related to its photon content when the field is quantized. The Fourier decomposition can be expressed in terms of helicity or circularly polarized basis which is convenient in order to calculate the helicity.

We have applied the Fourier method to compute a set of solutions such that at a particular time the field lines are torus knots. In the set it is included the Hopfion (which is a null field) and in general many non-null fields.

We believe that the calculations involved are at undergraduate level, so any student with a minimum background can rework and check them. Maxwell equations remains a fundamental cornerstone of our understanding of nature, but one might think that classical electromagnetic theory of light has obtained its limits of serviceability in term of fundamental research. We hope that this work proves the contrary. Solutions with new topological properties may pave a road for future discoveries.

\section*{Acknowledgements}
This work is supported by the research grant from the Spanish Ministry of Economy and Competitiveness ESP2017-86263-C4-3-R.

\end{document}